\documentclass[conference]{IEEEtran}
\usepackage{epsfig}
\usepackage{graphicx}
\usepackage{amsmath}
\usepackage{multirow}
\begin{document}
\title{ TCP-controlled Long File Transfer Throughput in Multirate 
WLANs with Nonzero Round Trip Propagation Delays }
\author{Pradeepa BK and Joy Kuri \\ 
Centre for Electronics Design and Technology, \\
Indian Institute of Science, Bangalore. India. \\
{bpradeep, kuri}@cedt.iisc.ernet.in
}
\markboth{}
{ \MakeLowercase{}}
\maketitle
\begin{abstract}
In a multirate WLAN with a single access point (AP) and several stations
 (STAs), we obtain analytical expressions for TCP-controlled long file 
 transfer throughputs \emph{allowing nonzero propagation delays between 
 the file server and STAs}. We extend our earlier work in 
 \cite{astn_model:pradeep_kuri} to obtain AP and STA throughputs in a 
 multirate WLAN, and use these in a closed BCMP queueing network model 
 to obtain TCP throughputs. Simulation show that our approach is able 
 to predict observed throughputs with a high degree of accuracy. 
\end{abstract}
\IEEEpeerreviewmaketitle
\section{Introduction}\label{sec:Introduction}
This paper is concerned with infrastructure mode WLANs that use IEEE 802.11 
DCF mechanism. We are interested in analytical models 
for evaluating  the performance of TCP-controlled downloads where
each link experiences propagation delay. 
A detailed analysis of the aggregate throughput of TCP flows in WLANs
for a single rate Access Point (AP) (where all stations (STAs) are associated with the AP
at a single rate) is given in \cite{astn_model:Kuriakose} by assuming 
negligible or zero round trip time (RTT).
Similarly, the performance of the AP is evaluated in the multi rate case in 
\cite{astn_model:Krusheel}, \cite{astn_model:pradeep_kuri} and 
\cite{astn_model:pradeep_kuri2}. 
However these works also ignore the RTT. Here in our work we model the AP by 
considering round trip propagation delay (RTPD) and hence RTT.

In this paper, we are interested in obtaining analytical expressions for 
TCP-controlled long file transfer throughputs in case nonzero RTTs. Clearly, 
this is the case that is most relevant in practice. In addition, we allow STAs
to be associated with the AP at one of a number of possible rates; for example,
in 892.11g, the rate association belongs to the set \{ 54, 48, 36, 24, 18,
12, 6 \} Mbps. Again this is common in practice, because STAs can be at varying 
distances from the AP.

We obtain the closed-form expressions and numerical evaluations  apply them in other contexts of practical relevance.  One such application, 
which we are working on now, is to utilize the results reported here in devising
an improved AP-STA association scheme. 

Our approach is divide the problem into two parts. The first part is to get the
model to represent the number of STAs with ACKs in MAC queues as an embedded 
Discrete Time Markov Chain (DTMC), embedded at the instants of successful 
transmission events. We consider a successful 
transmission from the AP as a reward. This leads to viewing the aggregate TCP
throughput in the framework of Renewal Reward theory as given for example in 
\cite{astn_model:Wolff}. We obtain expressions for network state 
probabilities, as well as the service rates of the AP and STAs. The second part
is to model the complete network, with nonzero RTT, as a closed BCMP queueing
network \cite{astn_model:bcmp}.

The main contribution of this paper is the analytical model for TCP-controlled
long file transfer throughput in a WLAN with nonzero RTPD, using a BCMP 
queueing network. Simulations indicate 
that download traffic scenario with RTPDs, our numerical
evaluation of analytical expression matches with error less than 3\% .

This paper is organized as follows: Section \ref{sec:Related_Work} outlines
related work. 
In Section \ref{sec:System_Model} we state the system model and we discuss
the assumptions in the modelling. 
In Section \ref{sec:Analysis} we obtain throughput analysis. 
In Section \ref{sec:Evaluation}, we present performance evaluation results. 
In Section \ref{sec:Conclusion} we present some key observations on the model,
and the results and we conclude the paper.
\section{Related Work}\label{sec:Related_Work} 
Numerous models and analyses have been proposed for wireless networks with 
TCP-controlled traffic, but very few consider propagation delays.
In \cite{astn_model:Leith}, RTT is considered in modelling the TCP traffic 
in a WLAN. However, the authors' interest was in showing that 802.11e supports
features that can be exploited to overcome certain TCP performance anomalies. 

\cite{astn_model:Kuriakose} and \cite{astn_model:Bruno4} provide a model for
single rate AP-STA WLAN assuming zero RTT and consider file transfers from
a server located in the LAN. An extension of this model in 
\cite{astn_model:Krusheel} considers two rates of association with long 
file uploads from STAs to a local server. The multirate case, with $k$ rates,
is analysed in \cite{astn_model:pradeep_kuri}. \cite{astn_model:pradeep_kuri2}
considers the single rate case, but allows simultaneous TCP uploads and
downloads with arbitrary maximum window sizes. 
\cite{astn_model:Bruno1} and \cite{astn_model:Bruno2} analyze TCP-controlled 
uploads and downloads in the presence of UDP traffic. However, the effect
of RTT on the network performance is ignored. 
The letter \cite{astn_model:Bruno3} gives the average value analysis of TCP 
performance with upload and download traffic without considering RTT.
In \cite{astn_model:Onkar}, finite buffer AP with TCP traffic in both upload and
download direction is analysed with delayed and undelayed ACK cases. They 
consider server system located on the Ethernet to which the AP is connected 
and hence number of packets ``in flight'' outside the WLAN is ignored. 

\cite{astn_model:Yu} provides an analysis for a given number of STAs and maximum
TCP receive window size by using the well known $p$ persistent model proposed
in \cite{astn_model:Cali}. However both \cite{astn_model:Yu} and 
\cite{astn_model:Cali} do not consider the effect of RTT on the performance.
In \cite{astn_model:Miorandi}, a queueing model is proposed to compute the mean
session delay of HTTP sessions in the presence of short-lived TCP flows and 
the impact of TCP maximum congestion window size on this delay is studied.
\section{System Model}\label{sec:System_Model}
\begin{figure}[ht]
\centering
\includegraphics[scale=0.5]{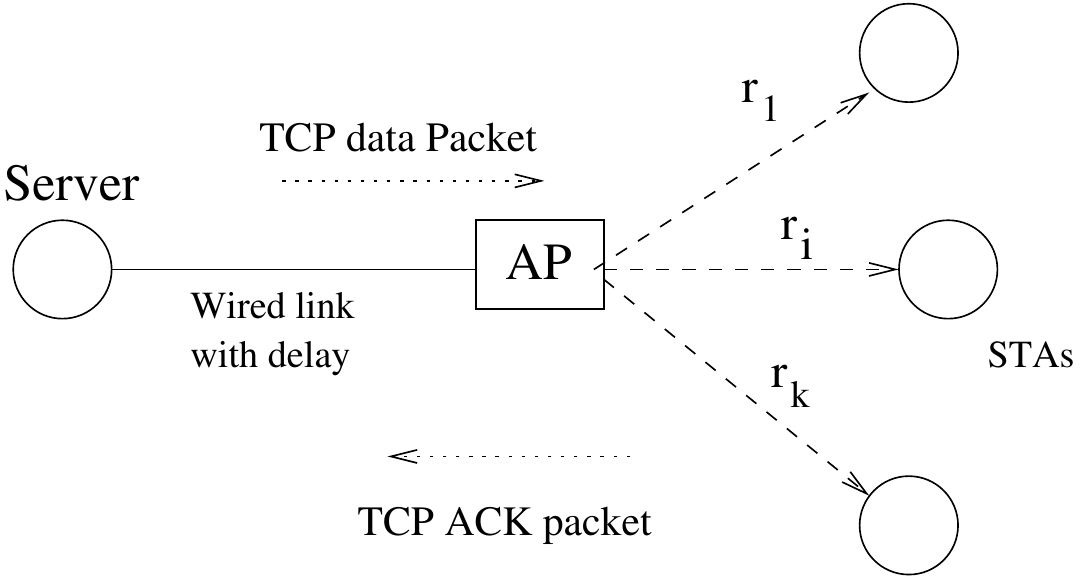} 
\caption{The network and traffic configurations. STAs are downloading 
long files from a server through an AP.}
\label{fig:AP_STAs}
\end{figure}

We consider a WLAN with $M$ STAs are associated to an AP as shown
in Figure \ref{fig:AP_STAs}.The STAs are downloading long files from a server
which is far away from the local wireless network. Hence there is a propagation
delay between the AP and the
server. Every packet experiences this delay. The AP sends TCP data packets to
these STAs. The arrows in Figure \ref{fig:AP_STAs} show the direction of
the data packets in the network. Since these are TCP links, there is also
feedback traffic composed of TCP-ACK packets.
 
 Every STA has a single TCP connection.Further, because of long file 
transfer scenario, we can assume that TCP
sources are operating in Congestion Avoidance mode. Hence TCP startup
transients can be ignored. TCP windows grow to the maximum value possible, 
i.e., the maximum receive window advertised by the receiver.  Also, TCP 
timeouts do not occur.

Both AP and STA contend for the channel using the DCF mechanism.
We assume that there are no link errors. Packets in the medium are 
lost only due to collisions. When the AP wins the channel, it delivers TCP
data packets towards the STA and the STA returns TCP-ACK packets again by
contending and winning the channel. Further, 
we assume that the AP uses the RTS-CTS mechanism while sending data packets, 
while the STAs use basic access to send ACK packets, which is more realistic 
and efficient as TCP-ACK packets are much shorted than TCP data packets. As 
soon as the STA receives a data packet, it generates an ACK packet without any 
delay and it is enqueued at the MAC layer for transmission. We assume that all 
the nodes have sufficiently large buffers, so that packets are not lost due to 
buffer overflow. These ACK packets travel through wired network and reach the 
server. Again server generates next window of TCP data packets. All packets 
from server experience the propagation delay, reach the WLAN and are enqueued 
at the AP to reach  STAs.
\section{Analysis}\label{sec:Analysis}
\subsection{AP and STA throughputs}
 Let $m_i $ be the number of STAs associated with the AP at 
the PHY rate $r_i$, where $i$ $ \in \lbrace 1,2, \ldots k \rbrace $ with $r_1  
>  r_2  > ... r_k $ as discussed in \cite{astn_model:Krusheel} and \cite
{astn_model:pradeep_kuri}. The probability that the AP sends a TCP data packet 
to an STA at rate $ r_i $ is $ p_i $. Part of the development (till 
Equation \eqref{eq:ap_thpt}) is along the lines of \cite
{astn_model:pradeep_kuri}; this is included here for completeness and readability.
\begin{figure}[h]
\centering
\includegraphics[scale=0.5]{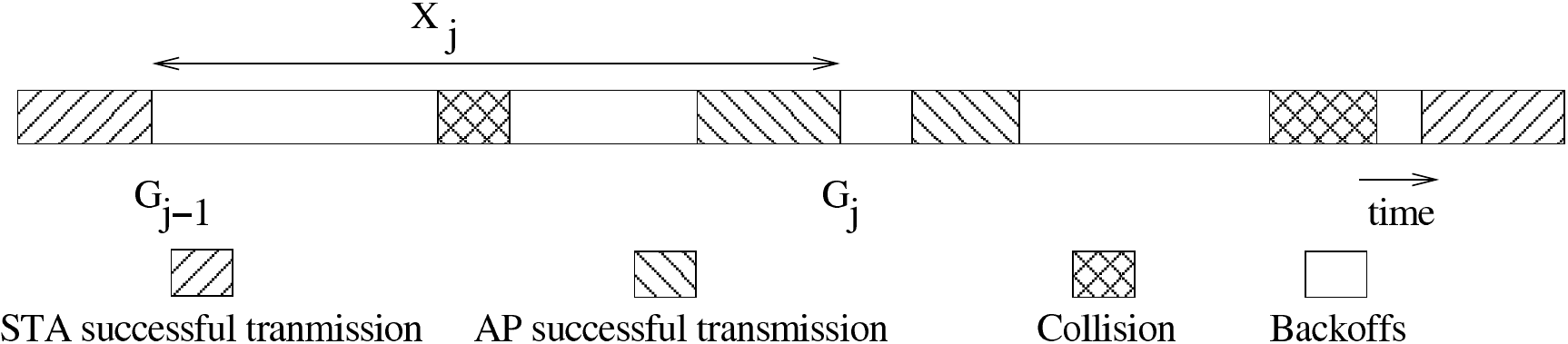} 
\caption{ Channel activity: $G_j$ are the random epochs at which successful 
transmissions end. Random variable $X_k$ denotes the duration of the $j^{th}$ 
contention cycle $[G_{j-1}, G_{j}) $. Each contention cycle consists of one or 
more back off and collisions slots but ends with a successful transmission. }
\label{fig:channel_activity}
\end{figure}
Consider Figure \ref{fig:channel_activity}, where a possible sample path of the 
events on the WLAN channel is shown. The random epochs $ G_j $ indicate the end 
of the $ j^{th} $ successful transmission from either the AP or one of the 
STAs. We begin by assuming that each $m_i$ is large. We observe that most 
STAs have empty MAC queues, because, in order for many STAs to have TCP-ACK 
packets, the AP must have had a long run of successes -- and this is unlikely 
because no special priority is given to the AP. So, when the AP succeeds in 
transmitting, the packet is likely to be for an STA with an empty MAC queue.

\begin{figure}
\centering
\includegraphics[scale=.4]{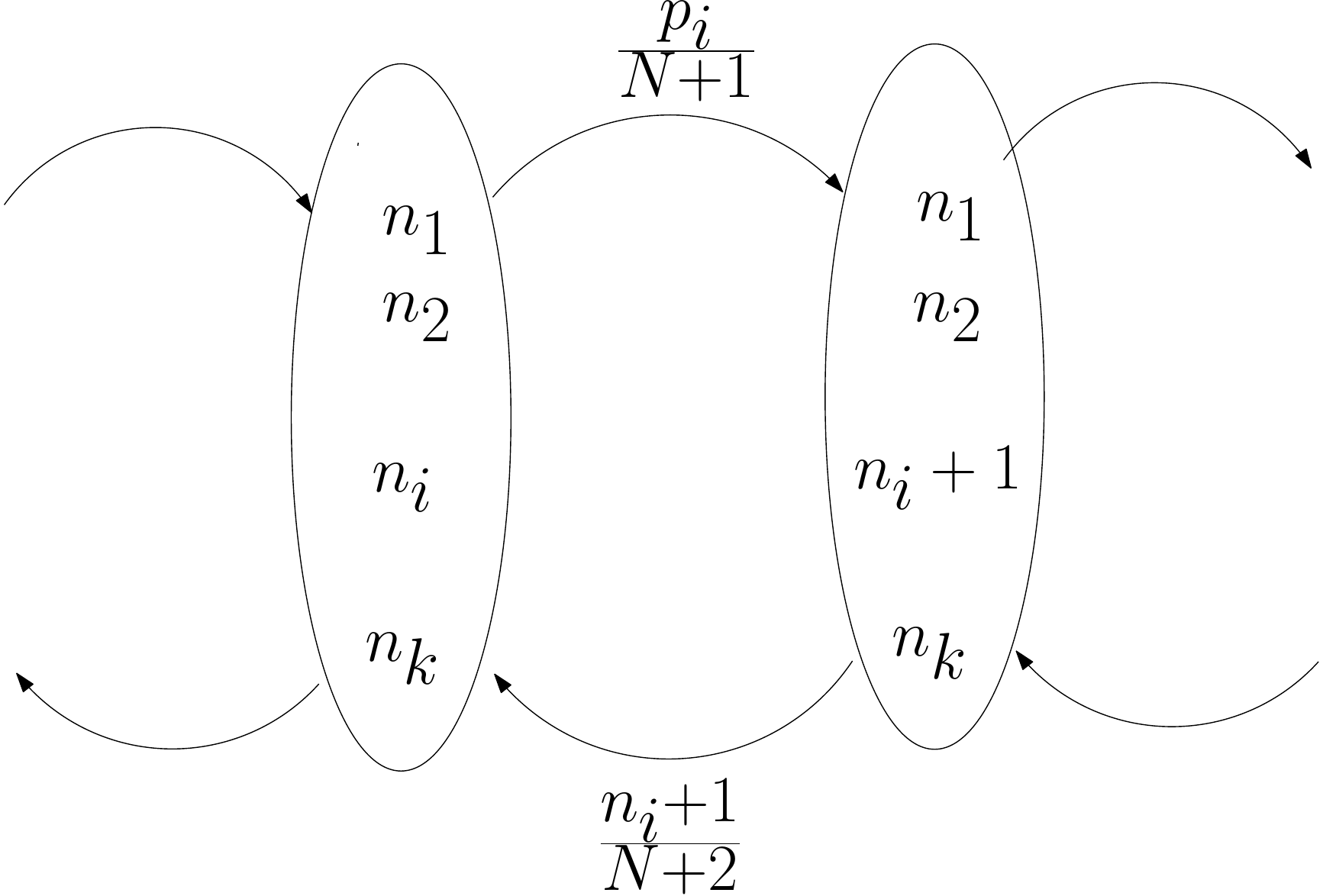}
\caption{Embedded Markov chain formed by the AP and $ n_1 + n_2 + ... + n_k = N 
$ stations associated with the AP at $ k $ different data rates }
\label{fig:MarkovChain}
\end{figure}

Let $ S_{i,j} $ be the number of STAs at rate $ r_i $, ready with an ACK. 
Let $ \sum_{i=1}^{k} S_{i,j} = N $ be the number of nonempty STAs. Since there
are $N$ nonempty STAs and a nonempty AP, each nonempty WLAN entity attempts to
transmit with probability $ \beta_{(N+1)} $ as in \cite{astn_model:kumar}. So 
$( S_{1,j}, S_{2,j},..., S_{k,j} )$ evolves as a Discrete Time Markov Chain
(DTMC) over the epochs $ G_j $. This allows us to consider
$((S_{1,j}, S_{2,j},..., S_{k,j} ) , G_j) $ as a Markov Renewal Sequence, and
$( S_{1}(t), S_{2}(t),..., S_{k}(t) )$ as a semi-Markov process.
We have a multidimensional DTMC which is shown in Figure \ref{fig:MarkovChain};
transition probabilities are indicated as well (we used in $n_i$ as running index). By inspection, we can say that 
the DTMC is irreducible. The Detailed Balance Equation holds for a properly
chosen set of equilibrium probabilities. The DBE is
\begin{equation}
\fontsize{8}{8} \selectfont
\pi(n_1,...n_i ,...n_k ) \frac{p_i}{(N+1)} 
= \pi(n_1,...(n_i +1),...n_k )\frac{(n_i+1)}{(N+2)} 
\label{eq:DBE}
\end{equation}
Here $ \pi(n_1,...,n_k ) $, $ n_1, n_2, ... n_k \in \lbrace 
0,1,2,...k \rbrace $ is the stationary distribution of the DTMC. From the set 
of equations given in \eqref{eq:DBE} and 
$ \sum _{n_1=0} ^{ \infty } \sum _{n_2=0} ^{ \infty }... \sum _{n_k=0} ^{ 
\infty } \pi(n_1,...,n_k )  = 1 $, the stationary distribution is
\begin{equation}
\pi(n_1,n_2,...n_i ,...n_k ) = (N+1) \Pi _{i=1} ^{k} \frac{(p_i)^{n_i} 
}{(n_i!)} * \frac{1}{(2e)}  
\label{eq:stationary_dist}
\end{equation}	
Let $ X $ be the sojourn time in a state $ ( S_{i,j}, ... S_{k,j} ) $. 
Conditioning on various events (idle slot, collision or successful 
transmission) that can happen in the next time slot, the
following expression for the mean cycle length can be written down: \\
\begin{equation}
\begin{split}
E_{n_1..n_k}X & = P_{idle}(\delta + E_{n_1..n_k}X)  + \Sigma_{i} ( P^{r_i} 
_{sAP}  T^{r_i} _{sAP} )\\
& + \Sigma_{i} ( P^{r_i} _{c} ( T^{r_i} _{c} + E_{n_1..n_k}X ) \\
& +\Sigma_{i} ( P^{r_i} _{sSTA}  T^{r_i} _{sSTA} ) \\
& +\Sigma_{i} \left( P^{r_i} _{cSTA} ( T^{r_i} _{cSTA} + E_{n_1..n_k}X) \right)  
\label{eq:enx11}
\end{split}
\end{equation}
In the above expression \eqref{eq:enx11}, $ P_{idle} $ is the probability of
the slot being idle. $ P^{r_i}_{sAP} $ is the probability that the AP wins the
contention and transmits the data packet with rate $ r_i$. $ P^{r_i}_{sSTA} $
is the probability that the STA wins the contention and transmits the data
packet with rate $ r_i$. Detailed expressions are tabulated in 
\cite{astn_model:pradeep_kuri}.
In the above expression, collisions correspond to different cases are as
follows. First, the third term in \eqref{eq:enx11} arises when the AP transmits
a TCP data packet to an STA at rate $ r_i $ and some other STAs are involved in
a collision. The second case (fifth term in \eqref{eq:enx11}) corresponds to an
STA transmitting a TCP ACK packet to the AP at rate $ r_i $ and some other node
transmitting simultaneously. In the above expression, various probabilities
have been obtained by considering the events and using channel access
probability $\beta _{N+ 1} $,when there are $(N + 1)$ contending nodes.

From Equation \eqref{eq:enx11} we have $ E_{n_1..n_k}X = $
\begin{equation}
\fontsize{6}{6} \selectfont
\frac{ P_{idle} + \sum P^{r_i} _{sAP} T^{r_i} _{sAP} + \sum 
P^{r_i} _{c}T^{r_i} _{c}  +\sum P^{r_i} _{sSTA} T^{r_i} _{sSTA} +\sum P^{r_i} 
_{cSTA} T^{r_i} _{cSTA}   }{1- P_{idle} -\sum P^{r_i} _{sAP} -\sum P^{r_i} _{c} 
-\sum P^{r_i} _{sSTA} -\sum P^{r_i} _{cSTA} }
\label{eq:enx}
\end{equation}
In Equation \ref{eq:enx}, calculations of probabilities and times are shown in \cite{astn_model:pradeep_kuri}.
We are interested in finding long run time average of successful transmissions
from the AP. This leads to Markov regenerative analysis or the renewal reward
theorem approach. To get mean renewal cycle length, we can use the mean sojourn
time given in Equation \eqref{eq:enx}. The mean reward in a cycle can be obtained
as follows. 
A reward of 1 is earned when the AP transmits a TCP data packet successfully by 
winning the channel. The probability of the AP winning the channel is  
$ \frac{1}{(n_1 + n_2 + ... n_k + 1)}$. Hence, the semi Markov 
process exits the state $ (n_1,n_2,... n_k )$ with probability $ \frac{1}{(n_1 
+ n_2 + ... n_k + 1)}$. A reward of 0 is earned with the probability $ ( 1- 
\frac{1}{(n_1 + n_2+ ... n_k + 1)} )$. Therefore, the expected reward is  $ 
\frac{1}{(n_1 + n_2 + ... n_k + 1)}$.

Hence, the aggregate TCP throughput for the AP in this case can be 
calculated as\\
\begin{equation}	
 \Phi_{AP-TCP} = \frac{\Sigma^{\infty}_{n_1 =0} \Sigma^{\infty}_{n_2 =0} ... 
 \Sigma^{\infty}_{n_k =0} \pi (n_1... n_k)\frac{1}{n_1+...+n_k} 
}{\Sigma^{\infty}_{n_1 =0} \Sigma^{\infty}_{n_2 =0} ... \Sigma^{\infty}_{n_k 
=0}  \pi (n_1... n_k)  E_{n_1,..n_k}X}  
\label{eq:ap_thpt}
\end{equation}
We are also interested in finding the mean TCP throughput for the STAs. A
reward of 1 is counted when any STA transmits a TCP-ACK packet successfully by
winning the contention. The probability of STA at rate $r_i$ winning the
contention is $ \frac{n_i}{(n_1 + n_2 + ... n_k + 1)}$. A reward of 0 is
counted with probability $ ( 1- \frac{n_i}{(n_1 + n_2+ ... n_k + 1)} )$.
Hence the TCP throughput for the STA at rate $r_i$ is\\
\begin{equation}	
 \Phi_{STA-TCP-(r_i)} = \frac{\Sigma^{\infty}_{n_1 =0} \Sigma^{\infty}_{n_2 =0} ... 
 \Sigma^{\infty}_{n_k =0} \pi (n_1... n_k)\frac{n_i}{n_1+...+n_k} 
}{\Sigma^{\infty}_{n_1 =0} \Sigma^{\infty}_{n_2 =0} ... \Sigma^{\infty}_{n_k 
=0}  \pi (n_1... n_k)  E_{n_1,..n_k}X}  
\label{eq:sta_thpt}
\end{equation}

\subsection{BCMP model}
\begin{figure}
\centering
\includegraphics[scale=0.5]{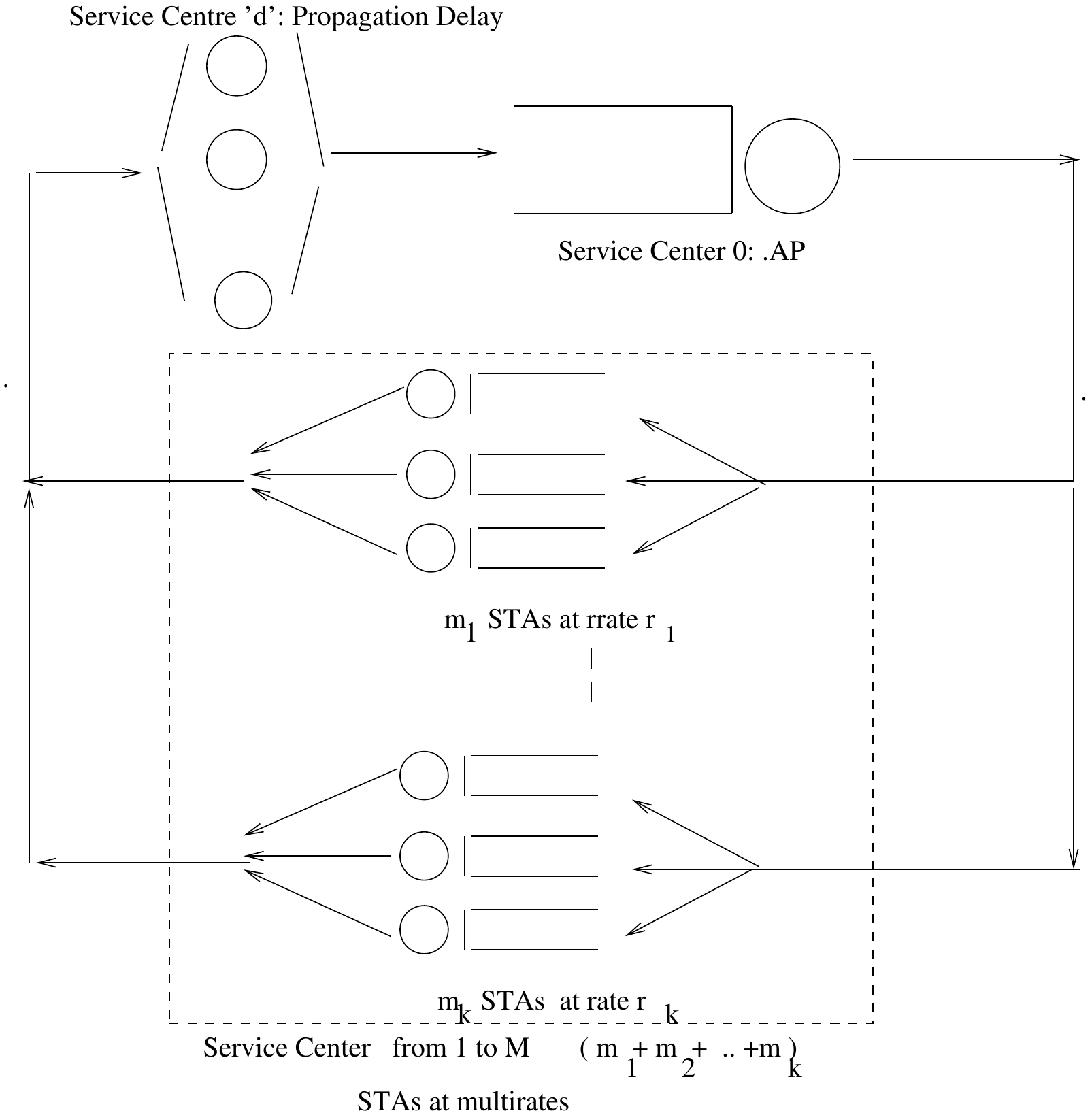} 
\caption{An equivalent queueing network model for the scenario shown in 
Figure \ref{fig:AP_STAs}, considering packets as customers. Total number
 of customers is the sum of maximum receive windows advertised by each receiver.}
\label{fig:threeQs}
\end{figure}

We can model the scenario shown in Figure \ref{fig:AP_STAs} as a BCMP closed
queueing network \cite{astn_model:bcmp} with service centers as shown in Figure 
\ref{fig:threeQs}. We consider RTPD as a delay center. Once wireless specific 
aspects are captured in $ \Phi_{AP-TCP}$ and $\Phi_{STA}$,
we consider the BCMP network service centres as if they were linked
by regular wired links. This is a modelling assumption for tractability.

Let us consider $ W $ packets in this network. The queues in this network 
representing the AP and STA are first come first served queues (FCFS) which 
are ``Type 1'' service centres in the terminology of \cite{astn_model:bcmp}. 
Similarly, the queue representing round trip propagation delay (RTPD) is an 
infinite server queue with deterministic service time, which is a ``Type 3'' service center.

Let the service rate of the AP be $ \tau $. Let us consider $ w_0 $ packets to
be at center 0. That is, $ w_0 $ among $ W $ packets are in the AP. Also, let 
$ w_1 $ out of $ W $ packets be in center 1, which is an STA at rate $r_1$ .
Similarly, $ w_i $ packets in center correspond to STA $i$. The remaining 
packets we say $w_d$ are in the delay center. The state of the network 
can be represented by $ S = ( x_0, x_1, x_2, x_3,..., x_M, x_d ) $,
as in \cite{astn_model:bcmp}. The definitions of $ x_0, x_1, x_2 ...$ depend on the type of the service center $i$  and are given in \cite{astn_model:bcmp}.

Let there be $m_j$ STAs at rate $r_j$. STAs at a particular rate constitute
 customers of a particular class in the BCMP network. We have $m_1 +m_2 + m_k = M$.

Every transition is both a departure from one center and an arrival at another 
center. For every $i$, let $ e_{i,j} $ be the fraction of transitions that are
arrivals at (departures from) center $i$. Let $ v_{i',j',i,j}$  be the 
probability that a customer of class $j'$ at center $i'$  goes to  center $i$ 
and becomes a class $j$ customer. From \cite{astn_model:Wolff}, $ e_{i,j} $  is the
unique solution (that sums to 1) of the following system of equations:
$ e_{i,j} = \sum_{i',j'} (e_{i',j'}) v_{i',j',i,j} $. 
From Figure \ref{fig:threeQs}, all the packets from the RTPD delay center go
to the AP: $v_{d,j,0,j} = 1$. The probability that a $j$ class customer from the
AP arrives to the $i^{th}$ service center is $ v_{0,j,i,j} = 1/(m_i)$ when $j -i $, and is 0 otherwise. 
The probability that a $j^{th}$ class customer from  the $i^{th}$ center arrives
to the delay center is $v_{i,j,d,j}= 1 $. The remaining probabilities are zero. 
It can be shown easily that for $k$ different possible rates,
$e_{0,j} = \frac{m_j}{3M} $ and $ e_{d,j} = \frac{m_j}{3M}$ for each $j$ and  the arrival rate into a class $j$ is $ \frac{ e_{0,j} }{m_j} = \frac{1}{3M} $.

By the BCMP theorem \cite{astn_model:bcmp}, the equilibrium probabilities
are given by 
\begin{equation}
\begin{split}
& P(S= x_0, x_1, x_2,..., x_M, x_d) \\
& = C d(S)f_0 (x_0) f_1 (x_1) f_2 (x_2)...f(x_M)f_d (x_d)
\end{split}
\label{eq:bcmp_dist}
\end{equation}
where $C$ is the normalizing constant chosen to make the equilibrium state probabilities sum to 1, $d(S)$ is a function of the number of customers in the system, and $f_i$ is a function that depends on the type of service center $i$. The exact form of the states $x_0, x_1, ... $ is shown in \cite{astn_model:bcmp}.

 Let us take $n_i$ as  the total population at service center $i$,
From \cite{astn_model:bcmp}, for the FCFS server, AP (center 0),
\begin{equation}
f_0(x_0) =  \left( \frac{1}{\tau} \right) ^{n_0} \Pi _{j=1}^{n_0} e_{0,x_{0,j} }
\end{equation}
where $ x_{0,j} $ represents the class of the $j^{th}$ customer in FCFS order at service center 0,
for the FCFS servers at STAs, for all $ i \in \lbrace  1,.., M \rbrace $
\begin{equation}
f_i(x_i) = \left( \frac{1}{ \mu_{i}} \right) ^{n_{i}} \Pi _{j=1}^{n_0} e_{i,x_{i,j} }
\end{equation}
where $x_{i,j}$ indicates the class of the $j^{th}$ customer in FCFS order at service center $i$,
and for the infinite server, delay model, center `d', is represented by
cascading of $c$ number of exponential servers in $c$ stages with service rate
$ \frac{1}{c \times t_{RTPD}} $  method (by considering large value of $c$) gives 
\begin{equation}
f_d(x_d) =  \Pi _{j=1}^{k} \Pi _{l=1}^{c} \left( \frac{ e_{d,j}}{ c \times \tau_{RTPD} } \right) ^{n_{d,j,l} } (1/n_{d,j,l}!)
\end{equation}
where $ n_{d,j,l} $ represents the number of class $j$ customers in stage $l$
of service at center $d$. 
For a closed network, $ d(S) = 1 $.
 The average number of  packets in the AP,  $ n_{AP } $, the average number of
packets in STA $i$,  $ n_{STA-i} $, and the average number of packets in 
propagation $ n_{RTPD}$ can be obtained by finding the marginal distributions from \eqref{eq:bcmp_dist}. 
 
From Figure \ref{fig:threeQs}, it is clear that  
\begin{equation*}
 n_{AP}  + \sum_{i=1}^{M} n_{STA-i} + n_{RTPD}  = W  \\
\end{equation*}
Let the throughput in the closed network of Figure \ref{fig:threeQs} be $t_H$.
Then, applying Little's Theorem to service center `d', we have
\begin{equation}
n_{RTPD} = t_H \times t_{RTPD} 
\end{equation}
\section{Evaluation}\label{sec:Evaluation}
To verify the accuracy of  the model, we performed experiments using the
Qualnet 4.5 network simulator \cite{astn_model:Qualnet}, with the IEEE 802.11b
standard. We take 2 STAs associated  at rate 5.5 Mbps and 3 STAs at rate
11 Mbps with control packets transmission rate at 2 Mbps.
RTPD is varied from 10ms to 90ms in steps of 10ms. TCP Receive window is taken
as 60 packets per link. In Table \ref{table:AP_buffer}, results are given with 95 \% confidence interval over 30 runs.
\begin{table}[ht]
\fontsize{8}{8} \selectfont
\centering 
\begin{tabular}{|c|c|c|c|c|}
\hline
& Analysis & \multicolumn{3}{|c|}{Simulation} \\ \hline
RTPD(ms) & Packets &	Mean 	&	Max 	&	Min \\
\hline
10		& 	297.9	&	296.862 		&	299.115 		&	294.608 \\
20 		&	295.2	&	294.181 		&	296.385 		&	291.977 \\
30 		&	292.5	&	291.425 		&	293.615 		&	289.235 \\
40		&	289.8	&	288.716 		&	290.899 		&	286.532 \\
50 		&	287.0	&	286.034 		&	288.23  		&	283.837 \\
60 		&	284.3	&	283.277 		&	285.389 		&	281.165 \\
70 		&	281.5	&	280.077 		&	282.383 		&	277.771 \\
80 		&	278.7	&	277.645 		&	279.74  		&	275.551 \\
90 		&	276.6	&	275.484 		&	277.528 		&	273.441 \\
\hline 
\end{tabular}
\caption{Number of packets in AP buffer for different values of RTPD.} 
\label{table:AP_buffer} 
\end{table}
\begin{table}[ht]
\fontsize{8}{8} \selectfont
\centering 
\begin{tabular}{|c|c|c|c|c|}
\hline
& Analysis & \multicolumn{3}{|c|}{Simulation} \\ \hline
RTPD(ms) & Packets &	Mean 	&	Max 	&	Min \\ 
\hline
10	&	2.58	&	2.718	&	2.765	&	2.672	\\
20	&	4.27	&	5.374	&	5.517	&	5.231	\\
30	&	7.16	&	7.862	&	8.061	&	7.663	\\
40	&	10.72	&	10.782	&	11.078	&	10.485	\\
50	&	13.18	&	13.778	&	14.05	&	13.505	\\
60	&	16.15	&	16.467	&	16.724	&	16.211	\\
70	&	18.26	&	19.083	&	19.475	&	18.691	\\
80	&	20.13	&	21.459	&	21.799	&	21.119	\\
90	&	23.59	&	24.768	&	25.252	&	24.285	\\
\hline 
\end{tabular}
\caption{Number of packets in ``in flight'' for different values of RTT.  } 
\label{table:RTT_buffer} 
\end{table}
\begin{table}[ht]
\fontsize{8}{8} \selectfont
\centering 
\begin{tabular}{|c|c|c|c|c|}
\hline
& Analysis & \multicolumn{3}{|c|}{Simulation} \\ \hline
RTPD(ms) & Packets &	Mean 	&	Max 	&	Min \\ 
\hline
10	&	0.12		&	0.128	&	0.132	&	0.124	\\
20	&	0.125	&	0.131	&	0.136	&	0.125	\\
30	&	0.121	&	0.127	&	0.131	&	0.123	\\
40	&	0.121	&	0.127	&	0.13	&	0.124	\\
50	&	0.122	&	0.129	&	0.134	&	0.124	\\
60	&	0.121	&	0.126	&	0.132	&	0.121	\\
70	&	0.12		&	0.125	&	0.129	&	0.121	\\
80	&	0.121	&	0.127	&	0.13	&	0.123	\\
90	&	0.128	&	0.13	&	0.134	&	0.127	\\
\hline 
\end{tabular}
\caption{Number of packets in STAs buffer at rate 11 Mbps and 5.5 Mbps for different values of RTT.} 
\label{table:STA_buffer} 
\end{table}
\begin{table}[h]
\fontsize{8}{8} \selectfont
\centering 
\begin{tabular}{|c|c|c|c|c|}
\hline
& \multicolumn{2}{|c|}{AP Throughput (packets/s) } & \multicolumn{2}{|c|}{STA Throughput (packets/s)} \\ \hline
RTPD(ms) & Analysis &	Simulation  & Analysis &	Simulation \\
\hline
10	&	274.8	&	276.831	&	69.6	&	68.968	\\
20	&	271.5	&	273.513	&	69.4	&	68.407	\\
30	&	271.1	&	273.058	&	69.3	&	68.32	\\
40	&	269.5	&	271.01	&	69.9	&	69.229	\\
50	&	268.1	&	270.615	&	68.2	&	67.862	\\
60	&	270.8	&	270.577	&	68.7	&	67.767	\\
70	&	268.5	&	270.457	&	67.4	&	68.049	\\
80	&	267.2	&	268.099	&	66.1	&	68.817	\\
90	&	263.4	&	264.357	&	66.3	&	67.323	\\ \hline 
\end{tabular}
\caption{Average Throughput of the AP and the STAs at different RTPD values obtained by Analysis and simulation } 
\label{table:AP_STA_thpt} 
\end{table}
 \begin{figure}
\centering
\includegraphics[scale=0.7]{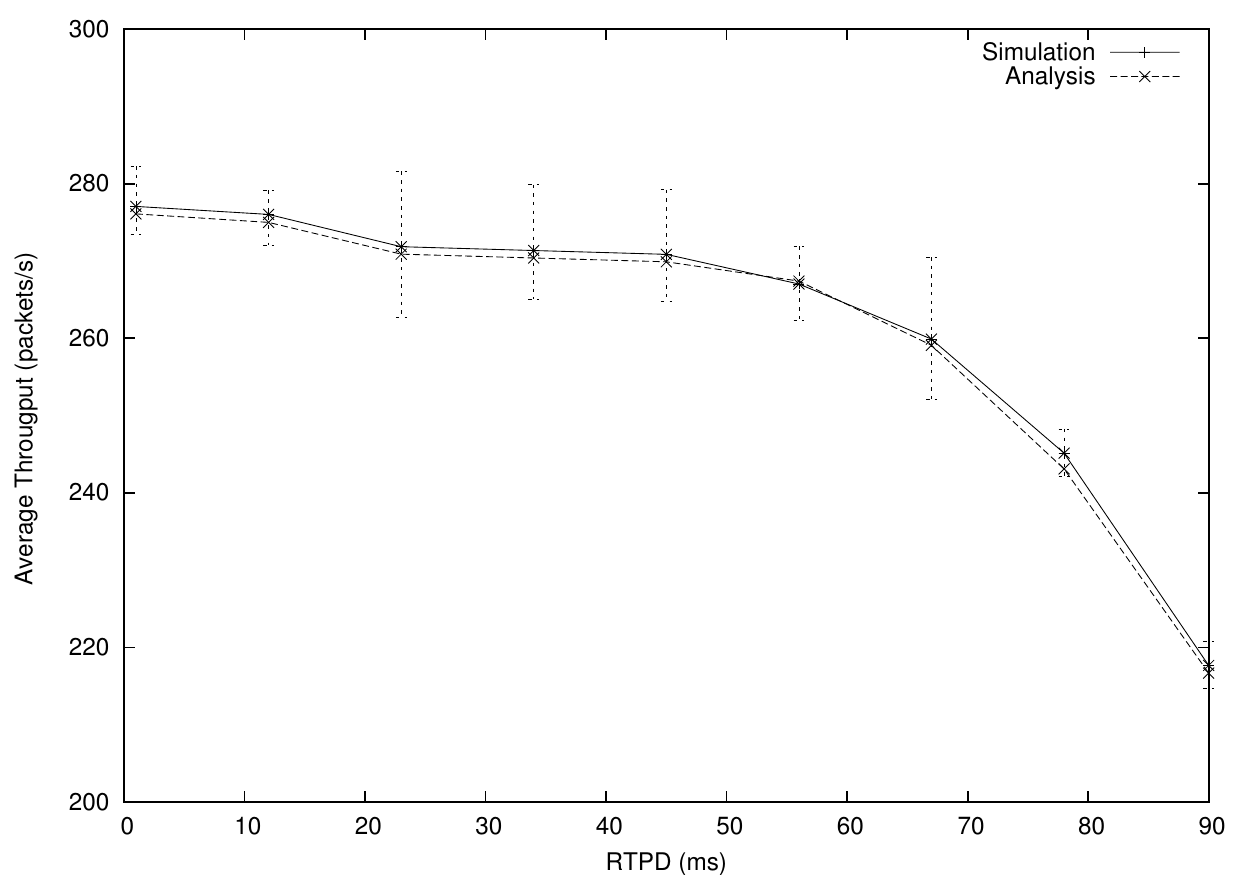} 
\caption{TCP throughput vs RTPD for TCP window of size 40 packets per link}
\label{fig:thpt}
\end{figure} 
 In Tables \ref{table:AP_buffer} to \ref{table:AP_STA_thpt} and Figure \ref{fig:thpt} comparisons between analytical and simulation values are given for selected data rates to illustrate the accuracy of the analytical model.
\section{Conclusion}\label{sec:Conclusion}
In this work, we presented an analytical model to obtain the aggregate
throughput for several TCP-controlled long file downloads in a network with
positive RTPD. We consider that TCP window sizes are the same for all 
connections to make the model simpler and to restrict our analysis to 
study the effect of RTPD and RTT on throughput. Our earlier work in 
\cite{astn_model:pradeep_kuri2} gives the effect of arbitrary TCP windows 
when RTPD is zero.

  In our simulation and numerical evaluation, we used the 802.11b standards.
However, our mathematical analysis is independent of  the parameters in these
standards. We can obtain similar analysis for other standards as well. We
assumed no packet losses; this is a topic for future work.  
\ifCLASSOPTIONcaptionsoff
  \newpage
\fi

%




\end{document}